\documentclass[]{interact}

\usepackage{epstopdf}% To incorporate .eps illustrations using PDFLaTeX, etc.
\usepackage[caption=false]{subfig}% Support for small, `sub' figures and tables

\usepackage[numbers,sort&compress]{natbib}% Citation support using natbib.sty
\bibpunct[, ]{[}{]}{,}{n}{,}{,}% Citation support using natbib.sty
\renewcommand\bibfont{\fontsize{10}{12}\selectfont}% Bibliography support using natbib.sty
\makeatletter% @ becomes a letter
\def\NAT@def@citea{\def\@citea{\NAT@separator}}% Suppress spaces between citations using natbib.sty
\makeatother% @ becomes a symbol again

\usepackage[english]{babel}
\usepackage{graphicx}
\usepackage{amsmath}
\usepackage{amstext}
\usepackage{amsfonts}%
\usepackage{dsfont}
\usepackage{float}
\usepackage{rotating}
\usepackage{xy}
\usepackage{pdflscape}
\usepackage{mathrsfs}
\usepackage{bm}
\usepackage{mathtools}
\usepackage{xcolor}
\newcommand{\indep}{\perp \!\!\! \perp}

\newcommand{\bs}{\mathbf{s}}

\newcommand{\bv}{\mathbf{v}}
\newcommand{\bx}{\mathbf{x}}
\newcommand{\bX}{\mathbf{X}}
\newcommand{\by}{\mathbf{y}}

\newcommand{\bSigma}{\bm{\Sigma}}
\newcommand{\bmu}{\bm{\mu}}
\newcommand{\given}{\,|\,}
\newcommand{\trans}{^\text{T}}

\newcommand{\cdf}{\emph{cdf}}
\newcommand{\pdf}{\emph{pdf}}

\DeclareMathOperator{\rank}{rank}

\theoremstyle{plain}% Theorem-like structures provided by amsthm.sty

\theoremstyle{definition}

\theoremstyle{remark}

\begin{document}

%\articletype{ARTICLE TEMPLATE}% Specify the article type or omit as appropriate

\title{A Vecchia Approximation for High-Dimensional Gaussian Cumulative Distribution Functions Arising from Spatial Data}

\author{
\name{Mauricio Nascimento\textsuperscript{a}\thanks{CONTACT Mauricio Nascimento. Email: mfn120@psu.edu} and Benjamin A. Shaby\textsuperscript{b}}
\affil{\textsuperscript{a}Department of Statistics, The Pennsylvania State University, University Park, PA 16802, USA; \textsuperscript{b}Department of Statistics, Colorado State University, Fort Collins, CO 80523-1801, USA}
}

\maketitle

\begin{abstract}
We introduce an  approach to quickly and accurately approximate the cumulative distribution function of multivariate Gaussian distributions arising from spatial Gaussian processes. This approximation is trivially parallelizeable and simple to implement using standard software.  We demonstrate its accuracy and computational efficiency in a series of simulation experiments, and apply it to analyzing the joint tail of a large precipitation dataset using a recently-proposed scale mixture model for spatial extremes.  This dataset is many times larger than what was previously considered possible to fit using preferred inferential techniques.
\end{abstract}

\begin{keywords}
Gaussian process ; Scale mixture ; Spatial extremes
\end{keywords}

\section{Introduction} 

We introduce a trivially parallelizable approach to quickly and accurately approximate the cumulative distribution function (\cdf) of multivariate Gaussian distributions with highly structured covariance matrices, such as those arising from spatial Gaussian processes. The multivariate Gaussian distribution is by far the most widely used for modeling multivariate and spatial data. To a large degree, its near universal adoption is the result of its simplicity; it is concisely and intuitively parametrized by a mean vector and pairwise dependence in the form of a covariance matrix.  Prominent examples of its use include time series models like autoregressive and moving average models, which consider the joint distribution of the observations observed at discrete time points to be multivariate Gaussian, as well as geostatistics models, which consider spatially-indexed observations to be realizations of a Gaussian process, usually with a parsimoniously parametrized covariance structure.  Even multivariate models that do not assume Gaussian responses often represent dependence using some kind of latent multivariate Gaussian distribution.

In most situations, likelihood-based inference on popular models just requires calculation of the joint density of all observations.  The probability density function (\pdf) for a multivariate Gaussian random variable is
\begin{align}
  \label{eq:gaussian-pdf}
    f(\bx ; \bmu, \bSigma) \equiv \phi_{\bmu,\bSigma}(\bx)=(2\pi)^{-k/2} \det (\bSigma) ^{-1/2} \exp \Big[-\frac{1}{2} (\bx-\bmu)' \Sigma^{-1} (\bx-\bmu)\Big],
\end{align}
where $\bmu$ is the mean vector of length $D$ and $\bSigma$ is the $D \times D$ covariance matrix.

In principle, there is nothing difficult about calculating this density; it simply requires commonplace operations like calculating an exponent, matrix determinant, matrix multiplication, and matrix inversion.  However this is not an easy task in practice when the dimension $D$ is large. The complexity of calculating the determinant and inverse of a $D \times D$ matrix is typically $O(D^3)$ for algorithms in common use. This means that for large values of $D$, the calculation of the {\pdf} becomes prohibitive.

%There is a large literature on approximating the Gaussian {\pdf}  \eqref{eq:gaussian-pdf} when $D$ is large, particularly when the covariance matrix $\bSigma$ is highly structured \cite[e.g.][and references therein]{Vecchia,stein2004approximating,Cressie2008,Banerjee2008,Nychka2015,katzfuss2017general,Katzfuss2017,Heaton2019}.  Here we focus on the case of high dimensional Gaussian distributions arising from stationary Gaussian processes, but the principles we use apply more broadly.  We will follow the approach that \citet{Vecchia} used for the \emph{pdf} and use it to approximate the \cdf.

Computing the Gaussian \cdf, which is a much more difficult problem, has received much less attention.  The problem has increased in prominence recently with advances in spatial modeling of extreme events.  State-of-the-art approaches for spatial extremes like \citet{Wadsworth2014}, \citet{Thibaud2016}, \citet{deFondeville2018}, and \citet{Huser2019} all require high-dimensional Gaussian {\cdf}s for inference.  This turns out to be the dominant computational bottleneck, and all but \citet{deFondeville2018} restricted their analyses to fewer than 20 spatial locations because larger datasets are computationally intractable using widely-used techniques for computing the Gaussian \cdf.  In real-world spatial applications, one should expect to see many more spatial locations, and existing approaches are not equipped to handle datasets of even moderate size.

Multivariate Gaussian {\cdf}s appear in other contexts as well; for example the density of multivariate skewed Gaussian and $t$ random variables are functions of the multivariate Gaussian \cdf \citep{SN2006}.  Here, we will focus on the case of spatial extremes.  To make things concrete, we will use the example of the Gaussian scale mixture model from \cite{Huser2017}, although our computational strategy would work equally well in any context with highly-structured covariance matrices.

Most approaches to calculating multivariate Gaussian probabilities are intended for problems of small or moderate dimension.  \citet{Genz1992} proposed a transformation from the original integral over $\mathbb{R}^D$ to an integral over a unit hypercube. Transforming to a finite region then allows the use any standard numerical integration method. \citet{Genz2004} derived formulas to calculate bivariate and trivariate Gaussian {\cdf}s with high precision using Gauss-Legendre numerical integration. The calculations are fast and precise but do not apply in higher dimensions.  \citet{Miwa2003} proposed a two-stage recursive approach to estimate the Gaussian \cdf. Their approach does not scale to high dimensions because it requires a sum over a combinatorially exploding (in $D$) number of terms.

The most popular approach for approximating Gaussian {\cdf}s in moderate dimensions was proposed by \citet{Genz2009}. They describe the use of Monte Carlo (MC) and quasi-Monte Carlo (QM) methods to estimate the joint \cdf.  Their QM methods have smaller asymptotic errors than the MC versions, and hence are the more widely used.

More recently, \citet{Genton2018} sped up the \citet{Genz2009} QM algorithm by performing matrix computations with fast hierarchical matrix libraries \citep{Hackbusch2015}.  As a follow-up, \citet{Cao2019} combined hierarchical matrix computations with a blocking technique to further speed up computations.  These approaches are much faster than their predecessors and work for Gaussian random variables with arbitrary covariance structures. They lean heavily on linking to specialized libraries for matrix operations.  Our approach achieves speedups using a fundamentally different strategy, by specifically leveraging the properties of highly-structured covariance forms arising from, for example, time series or spatial data.  It requires no exotic software, and is trivially parallelizable using simple tools in \texttt{R}.

\section{A Vecchia Approximation for the Multivariate Gaussian Distribution Function}
\label{sec:Gaussian}

The multivariate Gaussian {\cdf} that we wish to calculate is simply the integral of the {\pdf} \eqref{eq:gaussian-pdf}, 

\begin{align}
\label{eq:int}
  P(\bX < \bx; \bmu, \bSigma) \equiv \Phi_{\bmu,\bSigma}(\bx) =\int_{-\infty}^{x_1}\dots\int_{-\infty}^{x_k}\phi_{\bmu,\bSigma}(\by)\,dy_1\dots dy_D.
\end{align}

To calculate the integral \eqref{eq:int}, one must resort to numerical techniques, as it is well-known that no closed form exists, even in a single dimension.  In high dimensions, numerical integration is very difficult simply due to geometry and the curse of dimensionality.  The difficulty is compounded in the case of the Gaussian {\cdf} because while the curse of dimensionality requires an exponentially (in $D$) increasing number of evaluations of the integrand, the cost of each evaluation of the integration itself grows as $D^3$.  We seek a technique that simultaneously 1) reduces the effective dimension of the integral and 2) reduces the dimension of the {\pdf} in the integrand.

% \begin{align}
% \label{eq:int}
%     P(\mathbf{X}<\mathbf{x})=\int\dots\int_{D}\phi(y_1,\dots,y_n)dy_1\dots dy_n
% \end{align}

% Numerical methods are a useful way to evaluate integrals however they require the calculations of the function $\phi$ multiple times. This means that it is still necessary to calculate the determinant and the inverse of the covariance matrix. This is not a problem that can be ignored since, just like the ARIMA and spatial models, cumulative Gaussian distributions are use and multiple applications. 

% On section \ref{sec:GSMM} we will be analyzing the precipitation over Europe using the Gaussian Scale Mixture Model proposed by \cite{Huser2016}. The authors used a high dimension Gaussian distributions to defined the extremal dependence between different locations. When the observations are above some threshold, the multivariate density function is used, however when the observations are below the threshold the cumulative multivariate density function is necessary. More details about this model will be given on section \ref{sec:GSMM}.

\subsection{Vecchia Approximation for the Gaussian \pdf}

\citet{Vecchia} introduced a way to approximate high-dimensional Gaussian {\pdf}s arising from spatial data, which is particularly amenable to modification for our purposes.  The starting point of the \citet{Vecchia} approximation is to write the joint density as a product of cascading conditional densities,
\begin{align}
    \label{eq:cascading-conditional}
    f(\bx)=f(x_1) \prod_{i=2}^D f(x_i \given \bx_{1:i-1}).
\end{align}
Here, $f(x_1)$ is the univariate Gaussian density with mean $\mu_1$ and variance $\Sigma_{11}$, and, for $i = 2, \ldots, k$, the conditional density $f(x_i \given \bx_{1:i-1})$ is the univariate Gaussian density with mean $\mu_i + \bSigma_{[i,1:i-1]}\bSigma_{[1:i-1,1:i-1]}^{-1}(\bx_{1:i-1}-\bmu_{1:i-1})$ and variance $\Sigma_{i,i}-\bSigma_{[i,1:i-1]}\bSigma_{[1:i-1,1:i-1]}^{-1}\bSigma_{[1:i-1,i]}$.
The leading terms in this product are fast to calculate, but for terms corresponding to large $i$, the computations are nearly as burdensome as those of the original representation \eqref{eq:gaussian-pdf}.

 %$X_1\sim N(\mu_1,\Sigma_{[1,1]})$ and, for $i = 1, \ldots, k$, $X_i \given X_{1:i-1}\sim N(\overline{\mu},\overline{\Sigma})$ where $\overline{\mu}=\mu_i+\Sigma_{[i,1:i-1]}\Sigma_{[1:i-1,1:i-1]}^{-1}(X_{1:i-1}-\mu_{1:i-1})$ and $\overline{\Sigma}=\Sigma_{[i,i]}-\Sigma_{[i,1:i-1]}\Sigma_{[1:i-1,1:i-1]}^{-1}\Sigma_{[1:i-1,i]}$. This helps in the beginning but when $i=k$ we would need to calculate the inverse of a $k-1$ by $k-1$ matrix, which is not easier than a $k$ by $k$.

To help solve this problem, \citet{Vecchia} proposed an approximation to the full joint distribution, in the setting where the random vector $\bX$ is observed from a spatial Gaussian process. He modified the cascading conditional representation \eqref{eq:cascading-conditional} by replacing the conditioning on high-dimensional vectors $\bx_{1:i-1}$ with conditioning on well-chosen vectors that have much smaller dimension.  By limiting the conditioning sets to vectors of length $m << D$, this strategy replaces expensive $\mathcal{O}(D^3)$ matrix operations with much faster $\mathcal{O}(m^3)$ matrix operations.  The approximation of the joint density is then
\begin{align}
    \label{eq:vecchia-pdf}
    f(\bx) \approx f(x_1) \prod_{i=2}^D f(x_i \given \bx_{\mathcal{N}_i}),
\end{align}
where $\mathcal{N}_i$ is the conditioning set of size $m$ (more precisely, $\min(m,i-1)$) chosen for the component $x_i$, for $i=2, \ldots, D$.

A good choice for a conditioning set to approximate the complete conditional density of each $x_i$ might be the $m$ components that are most correlated with $x_i$.  In the context where the random vector $\bX = (X(\bs_1), \ldots, X(\bs_k))\trans$ arises from a stationary spatial Gaussian process observed at locations $\bs_1, \ldots, \bs_k$, the components most correlated with $X_i \equiv X(\bs_i)$ will be those observed at locations that are the $m$ nearest neighbors to $\bs_i$ (under covariance models in common use).  Other strategies for constructing conditioning sets have also been explored \citep{stein2004approximating,Guinness2016}.

Vecchia's approximation has been found to be quite accurate under many covariance models and sampling scenarios relevant to analysis of spatial Gaussian processes \citep{Guinness2016}.  Moreover, it is very fast to compute, even using the most naive implementation.  However, its power is fully realized when the $D$ components of the product are computed in parallel, which is trivially easy to implement using standard tools in \texttt{R}.

\subsection{Extending the Vecchia Approximation for the Gaussian \cdf}

Our approach to approximating the high-dimensional Gaussian {\cdf} is to re-write the joint {\cdf} as a telescoping product of conditional {\cdf}s, analogously to \eqref{eq:cascading-conditional}, and then to approximate each complete conditional  {\cdf} with {\cdf} that conditions on a smaller collection of components, analogously to \eqref{eq:vecchia-pdf}.  In the case of the {\pdf}, this strategy of choosing smaller conditioning sets eliminates the need to compute high-dimensional matrix computations required by \eqref{eq:gaussian-pdf}, whereas in the case of the {\cdf}, this strategy eliminates the need to compute  the high-dimensional integral required by \eqref{eq:int}.

Specifically, we can re-write any joint {\cdf} as 
\begin{align}
    \label{eq:cascading-conditional-cdf}
    F(\bx)= P(\bX < \bx) & = P(X_1 < x_1) \prod_{i=2}^D P(X_i < x_i \given X_1 < x_1, \ldots, X_{i-1} < x_{i-1}) \nonumber \\
    & = P(X_1 < x_1) \prod_{i=2}^D P(X_i < x_i \given \bX_{1:i-1} < \bx_{1:i-1})
\end{align}
Then, just as in the approximation to the {\pdf} \eqref{eq:vecchia-pdf}, in the {\cdf} \eqref{eq:cascading-conditional-cdf} each conditional probability in the product can be approximated by reducing the size of the conditioning set to at most $m$ components.  Thus, our Vecchia approximation for the Gaussian {\cdf} is
\begin{align}
    \label{eq:vecchia-cdf}
    F(\bx)  & \approx P(X_1 < x_1) \prod_{i=2}^D P(X_i < x_i \given \bX_{\mathcal{N}_i} < \bx_{\mathcal{N}_i}) \nonumber \\
    & = P(X_1 < x_1) \prod_{i=2}^D \frac{P(X_i < x_i, \bX_{\mathcal{N}_i} < \bx_{\mathcal{N}_i})} {P(\bX_{\mathcal{N}_i} < \bx_{\mathcal{N}_i})} \nonumber \\
    & = \Phi(x_1) \prod_{i=2}^D 
        \frac{ \Phi(\bx_{\{i,\mathcal{N}_i\}})}{ \Phi(\bx_{\mathcal{N}_i})},
\end{align}
where again $\mathcal{N}_i$ is the conditioning set of size $\min(m,i-1)$ chosen for the component $x_i$, for $i=2, \ldots, D$.

The approximation given by \eqref{eq:vecchia-cdf} reduces computational costs by replacing the $D$-dimensional integral in \eqref{eq:int} with a series of much simpler integrals of dimension $m+1$ and $m$, for $m << D$. Furthermore, all of the elements in the product can be computed in parallel.

The multivariate {\cdf}s in \eqref{eq:vecchia-cdf} still have to be evaluated numerically.  For all but the smallest possible choices of $m$, best practices suggest using a QM method like that of \citet{Genz2009} to approximate the numerator and denominator.

Similarly to the original Vecchia approximation to the Gaussian {\pdf}, choosing the conditioning sets involves a trade-off; choose $m$ too small and the accuracy of the approximation will suffer, but choose $m$ too large and the computational benefits will diminish.

\section{Simulation Study}%Plots
\label{sec:Simulation}

\begin{figure}
    \centering
    \includegraphics[scale=0.75]{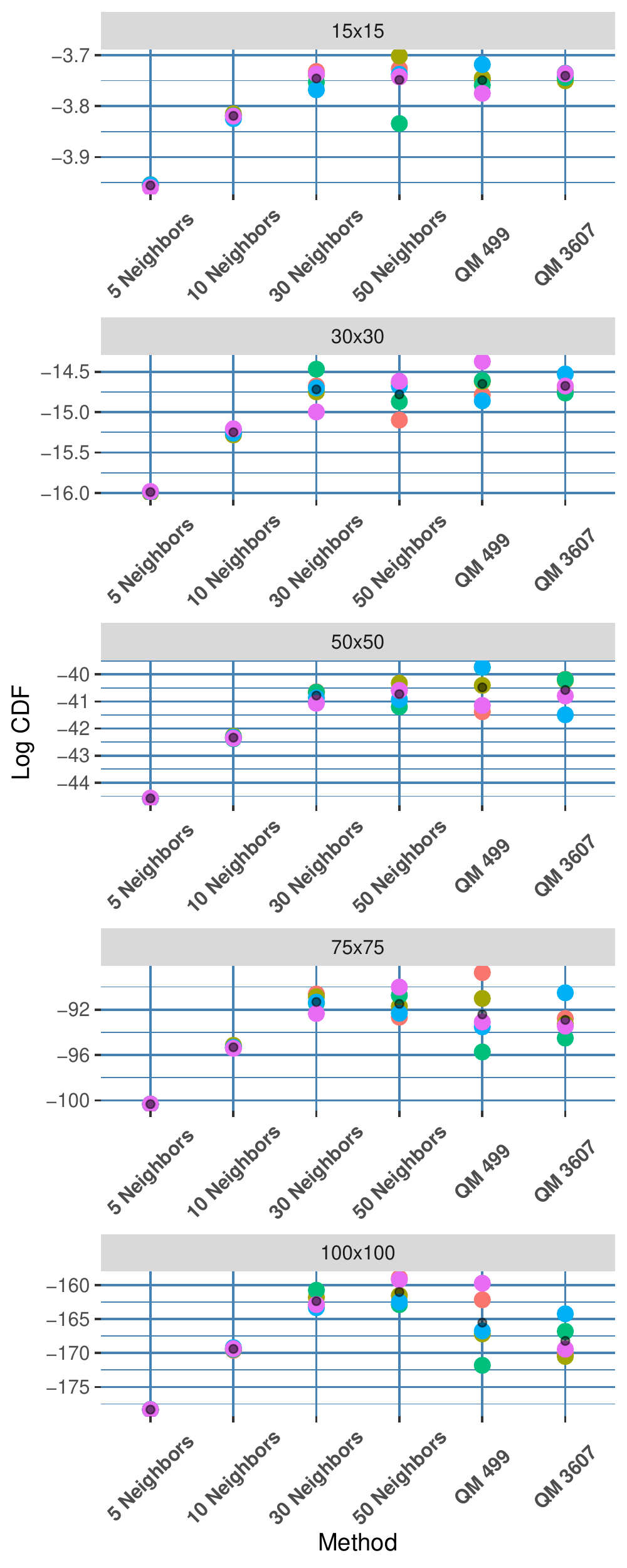}
    \caption{Estimated log {\cdf} for exponential Gaussian processes with range parameter $\rho=1$. The $x$-axis represents the different methods used for the {\cdf} computation and the $y$-axis is the log {\cdf}. Each point is an independent estimate of the log {\cdf}, and each black point is the average over the replications.  The Vecchia approximation seems to stabilize when at least 30 neighbors are used, and results in values that are consistent with the QM approximations.}
    \label{fig:SimRho1}
\end{figure}

\begin{figure}
    \centering
    \includegraphics[scale=0.75]{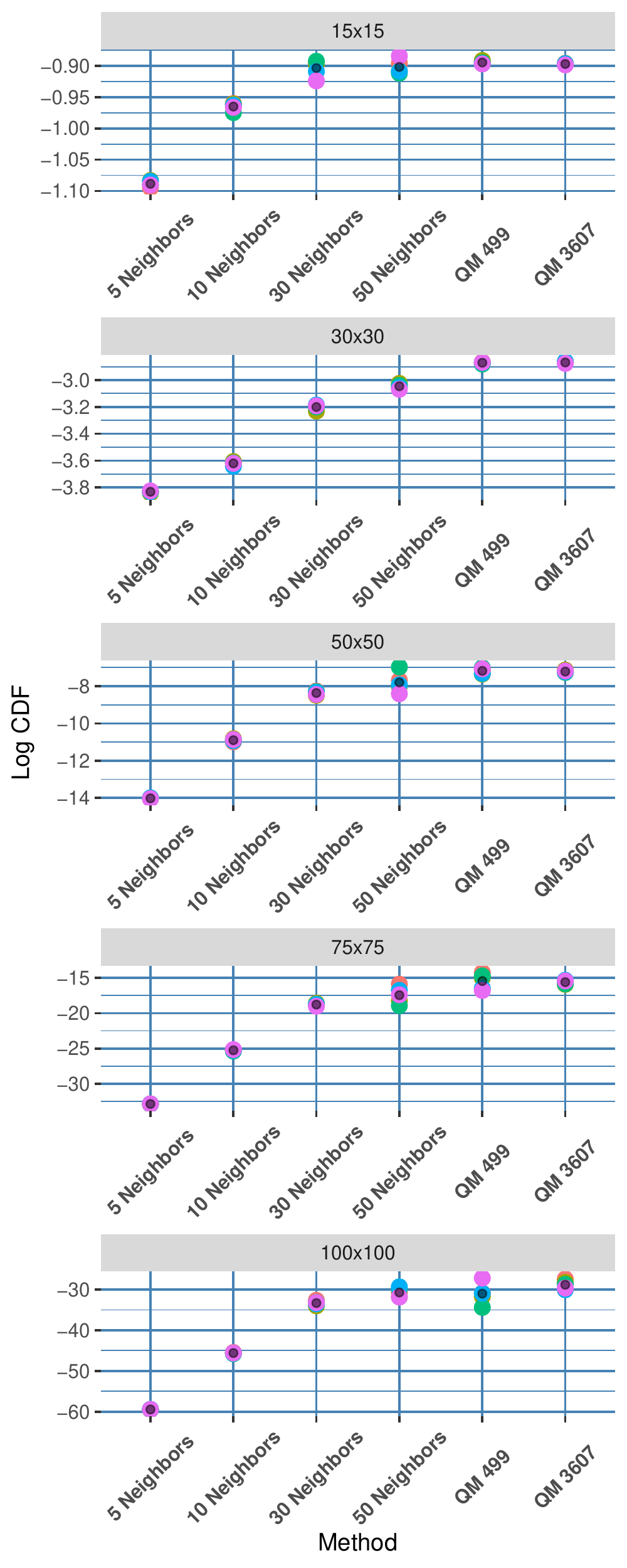}
    \caption{Estimated log {\cdf} for exponential Gaussian processes with range parameter $\rho=5$. The $x$-axis represents the different methods used for the {\cdf} computation and the $y$-axis is the log {\cdf}. Each point is an independent estimate of the log {\cdf}, and each black point is the average over the replications.  For this process with longer-range dependence, the Vecchia approximation may not stabilize until at least 50 neighbors are used, when results become consistent with the QM approximations.}
    \label{fig:SimRho5}
\end{figure}

\begin{figure}
    \centering
    \includegraphics[scale=0.8]{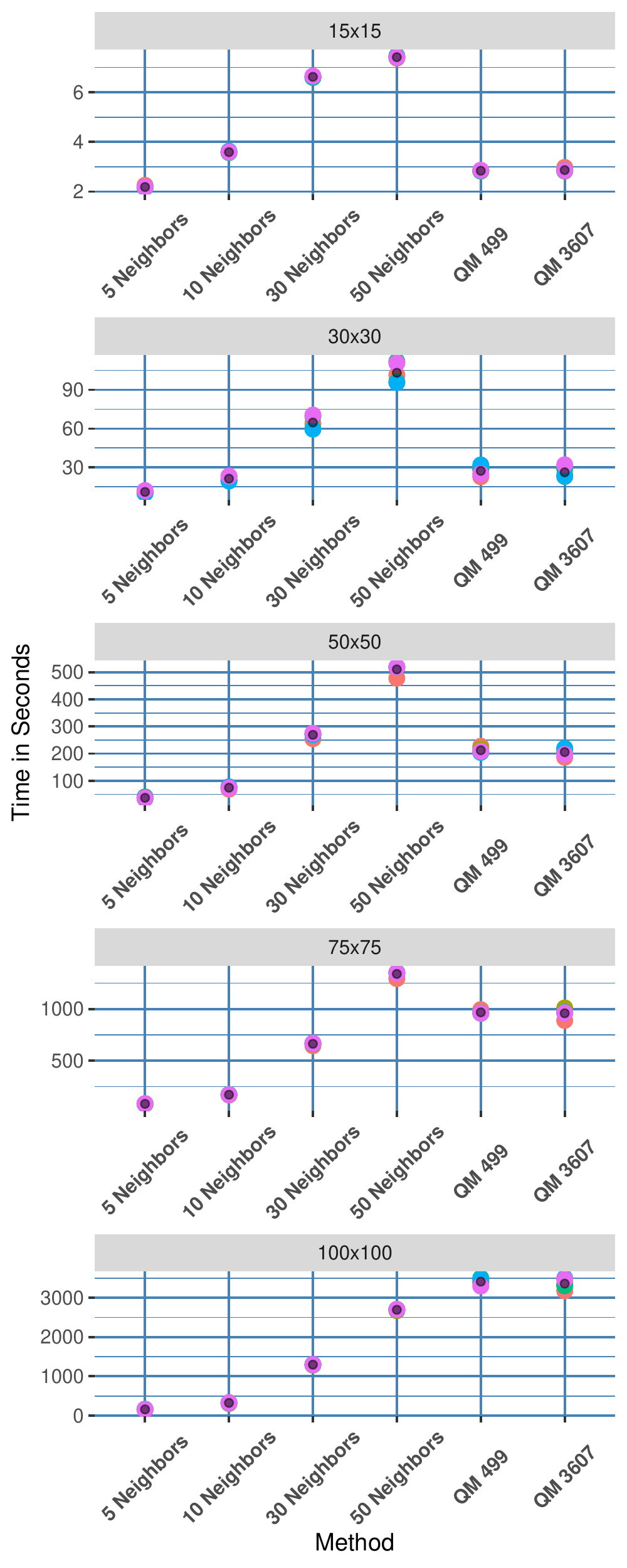}
    \caption{Time to estimate the {\cdf} approximation for an exponential Gaussian process with range parameter $\rho=1$. The $x$-axis represents the different approximation methods, and the $y$-axis is the computation time. Each point is an independent replication of the procedure, and the black point is the average over the replications.}
    \label{fig:TimeRho1}
\end{figure}

\begin{figure}
    \centering
    \includegraphics[scale=0.8]{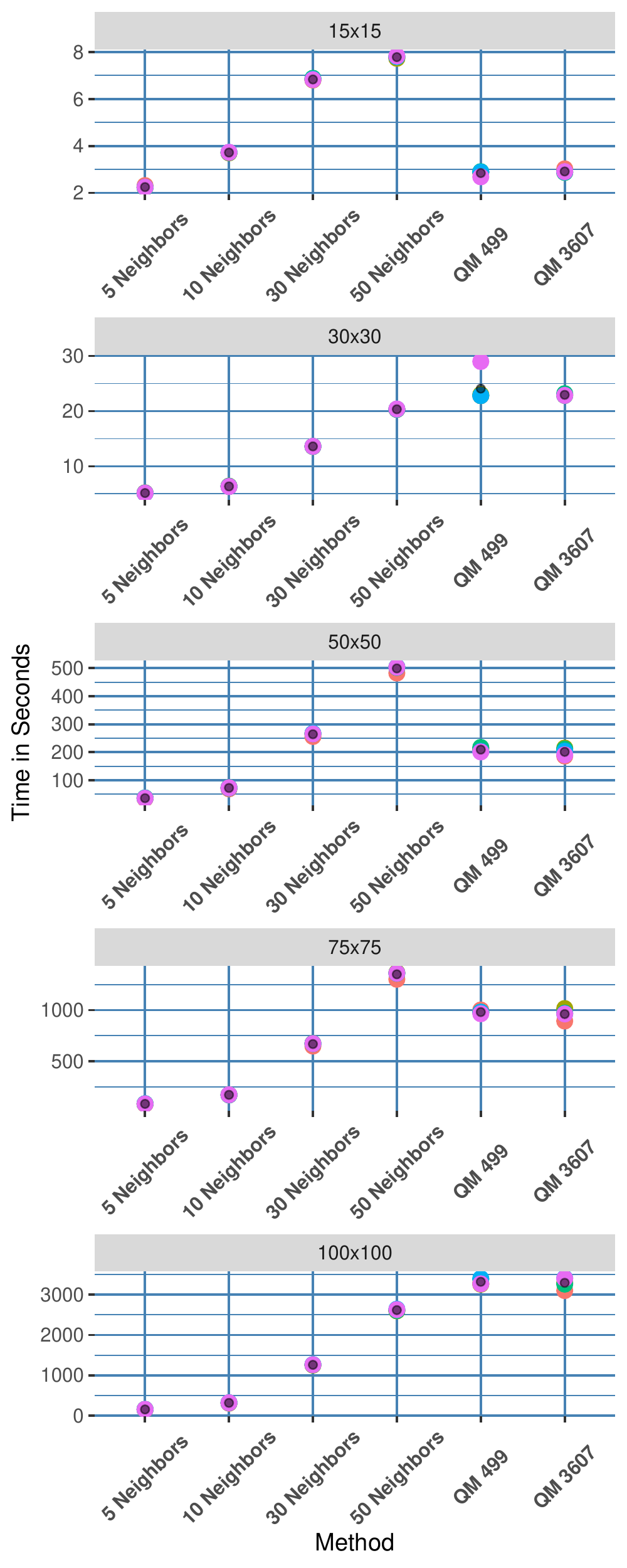}
    \caption{Time to estimate the {\cdf} approximation for an exponential Gaussian process with range parameter $\rho=5$. The $x$-axis represents the different approximation methods, and the $y$-axis is the computation time. Each point is an independent replication of the procedure, and the black point is the average over the replications.}
    \label{fig:TimeRho5}
\end{figure}

To assess the accuracy and speed of this approximation, and to explore the trade-off inherent in the choice of $m$, we conduct a simulation study.  Since the true value of the {\cdf} is not available, the best we can do to check for accuracy is to see whether it is consistent with results obtained from direct use of the \citet{Genz2009} QM approach.  We simulate a Gaussian process observed on equally spaced grids of five different sizes, $15\times15$, $30\times30$, $50\times50$, $75\times75$ and $100\times100$. We try two different covariance functions for the Gaussian process to see whether this has an impact on the {\cdf} estimation: an exponential model with range parameter $1$ and and exponential model with range parameter $5$, each with unit variance.  This makes a total of $10$ different scenarios. For each scenario, we used four different sizes of conditioning sets, choosing $m=$ $5$, $10$, $30$ and $50$ closest neighbors.  For comparison, we computed the \citet{Genz2009} QM method using 499 and 3,607 sample points.  We use the implementation of the \citet{Genz2009} algorithm in the \texttt{mvPot} package \citep{package-mvPot} for \texttt{R}. In principle, the accuracy and computational requirements of the QM grows with the number of sample points (which, here, must be a prime number).  Since the algorithms are stochastic, we repeated each calculation five times and plotted each replication as a dot in Figures \ref{fig:SimRho1}, \ref{fig:SimRho5}, \ref{fig:TimeRho1}, and \ref{fig:TimeRho5}.

Figure \ref{fig:SimRho1} shows the  value of the estimated log {\cdf} for all grid sizes and all estimation methods for the simulated Gaussian process with range parameter 1. The log {\cdf} estimated with the Vecchia approximation increases with the number of neighbors until it stabilizes for 30 neighbors, after which it is consistent with the two QM approximations. This suggests that, under this scenario, it is advisable to use at least 30 neighbors in order to estimate the log {\cdf}. For the two smaller grids, it appears that the Vecchia approximation has a similar variance to the QM approximation using 499 sample points, but a higher variance that the QM approximation using 3,607 sample points.  For the larger grids, the Vecchia approximation appears to have a lower variance than both QM approximations.  Figure \ref{fig:SimRho5} shows the same as Figure \ref{fig:SimRho1}, but for exponential Gaussian processes with range parameter 5.  The story is similar to the case with the shorter range process, except it appears that 50 neighbors may be necessary in order to stabilize the estimated log {\cdf}.  It may be the case that the number of neighbors necessary to accurately approximate the log {\cdf} increases with length of the dependence of the Gaussian process. Intuitively, this may occur because for processes with longer-range dependence, a smaller proportion of the information in data may be captured by local approximations.

Figures \ref{fig:TimeRho1} and \ref{fig:TimeRho5} show the time required to approximate the log {\cdf}, on a single core, for Gaussian processes with range parameters of 1 and 5, respectively. The computation time is influenced by both the number of observations and number of neighbors used in the Vecchia approximation. Computational costs increase with the number of observations, for both the Vecchia and QM approximation methods, and also increase with the number of neighbors in the conditioning set.  Oddly, the empirical computation time did not increase for the QM approximation with the larger set of sample points. For smaller grid sizes, the QM methods are faster than the Vecchia approximations, except when the size of the conditioning set very small.  For grids of size $50 \times 50$ and larger, computation time of the approximation using 30 neighbors was as fast as or faster than the QM method. When the number of observations is extremely large, in the case of the $100 \times 100$ grid, the computation time was much lower for the Vecchia approximation compared to the QM approximation. This suggests that for high-dimensional datasets the use of the Vecchia approximation is preferable QM method, even if computations are done sequentially.

\subsection{Parallel Computing}

\begin{figure}
    \centering
    \includegraphics[scale=0.6]{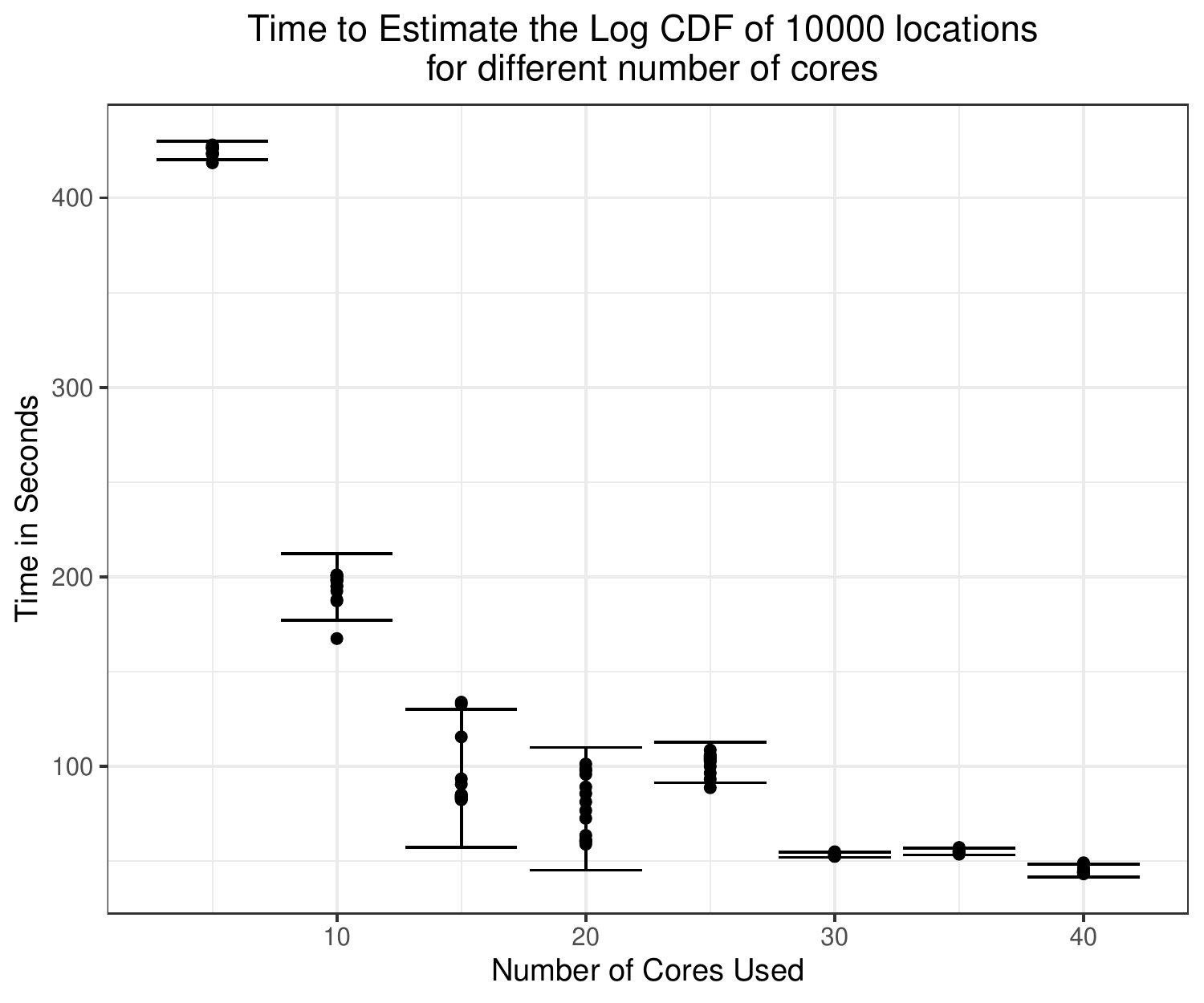}
    \caption{Time to compute log {\cdf} approximation parallelized across different numbers of computing cores.}
    \label{fig:TimeCores}
\end{figure}

Since each term of the Vecchia {\cdf} approximation \eqref{eq:vecchia-cdf} is independent of every term, it is trivial to parallelize the computations.  In practice, we compute all of the required low-dimensional Gaussian {\cdf}s on the log scale, and then sum them at the end.  In principal, the speedup should be linear in the number of cores used for the calculation. To explore this relationship, we compute the {\cdf} approximation based on a Gaussian process observed at 10,000 locations, varying the number of compute cores used between 5 and 40.  For each setup, we repeat the computation 15 times.  Figure \ref{fig:TimeCores} shows time required to compute the log {\cdf} approximation.  The computing time decreases with the number of cores. We observe roughly the expected linear relationship up to 20 cores, when a jump occurs before again decreasing.  We suspect that this is behavior a result of the particular hardware configuration we used, which consists of networked 20-core processors.  That is, we guess that once an additional physical processor is engaged, which occurs beyond 20 cores, overhead costs increase and attenuate the expected computational gains.  When 40 cores were used, it took less than 1 minute to compute the log {\cdf} approximation for 10,000 observations.  There are clearly some diminishing returns due to communication overhead, but in principle, this approximation could be made arbitrarily fast with a big enough computing system.

% Figure \ref{fig:LogCumLikCores} shows the estimated log CDF using different number of cores, we can see that the value of the log CDF does not depend on the number of cores. Figure \ref{fig:TimeCores} is the time to estimate the log CDF, the estimation time decreases with the number of cores used to calculate. When 40 cores were used calculate the probability it took less than 1 minute to estimate the log CDF of 10000 observations.

% \begin{figure}
%     \centering
%     \includegraphics[scale=0.6]{figure/Parallel/002Joint/density.pdf}
%     \caption{Estimated log CDF using different number of cores.}
%     \label{fig:LogCumLikCores}
% \end{figure}

\subsection{Effect of Neighbor Selection and Joint Estimation}

The representation defined by equation \eqref{eq:cascading-conditional-cdf} and its approximation \eqref{eq:vecchia-cdf} calculates the joint probability as the product of univariate conditional distributions.  However it is also possible to write the full joint {\cdf} as a cascading product of multivariate, rather than univariate, conditional {\cdf}s.   Under equation \ref{eq:vecchia-cdf}, it is necessary to calculate the $n$ univariate conditional probabilities, each of which requires a  $m+1$-dimensional {\cdf} calculation.  If instead we divide the components into $q$ groups of $p$ joint observations, such that $q\times p=n$, we would only need to calculate the product of $q$ conditional probabilities.  However, doing so would make the dimensionality of each individual Gaussian {\cdf} calculation in \eqref{eq:vecchia-cdf} between $m+p$ and $pm+p$.  So it would trade the cost of computing higher-dimensional {\cdf} terms for the benefit of computing fewer terms.  Such a trade-off could affect both the accuracy and computational efficiency of the approximation. \citet{Guinness2016} explored this possibility in the context of {\pdf}s and found that it can be advantageous to consider multivariate conditional densities in the Vecchia density approximation. To explore the effect of calculating higher dimensional conditional probabilities, we calculate the log {\cdf} approximation based on groupings of observations of different sizes. 

An additional consideration that could effect the accuracy and speed of the approximation is the construction of the conditioning sets.  Using the nearest neighbors, as we have done above, requires the additional step of ordering the components by distance, which could be slow.  Choosing randomly-selected conditioning sets could potentially speed up the computation by avoiding this sorting step. % We analyzed 2 different methods of selecting conditioning sets. The  first method uses of the most correlated, nearest neighbors when the observations are spatially located, observations to each individual observation to estimate the conditional probabilities. The second method correspond to use random observations that are not necessarily the most correlated. 

\begin{figure}
    \centering
    \includegraphics[scale=0.6]{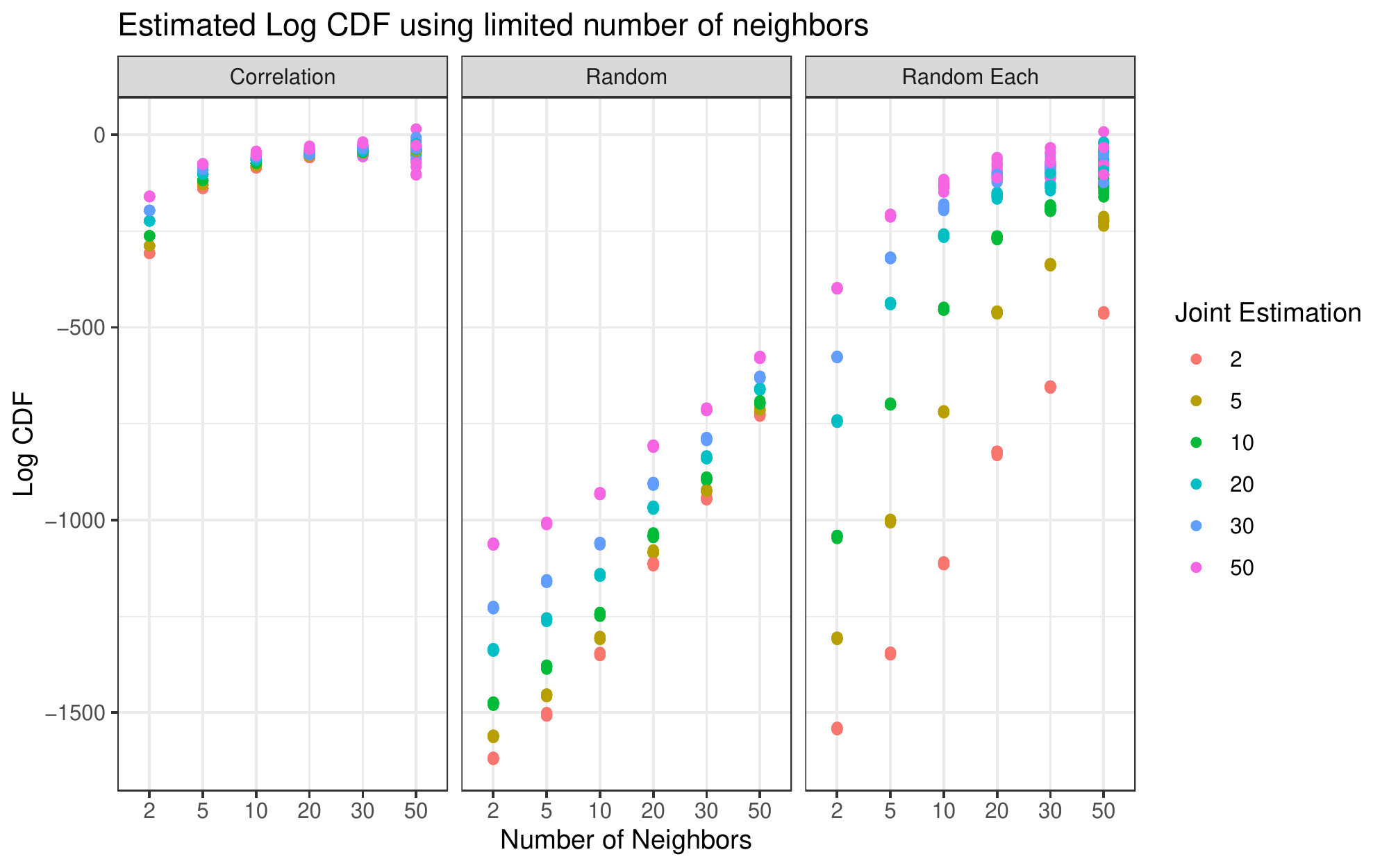}
    \caption{Estimated log {\cdf} based on observations from an exponential Gaussian process with range parameter $\rho=1$, using 3 different methods to select conditioning sets, and different dimensionalities of joint conditional observations.}
    \label{fig:LogCumLikNeighborsJoint}
\end{figure}

\begin{figure}
    \centering
    \includegraphics[scale=0.6]{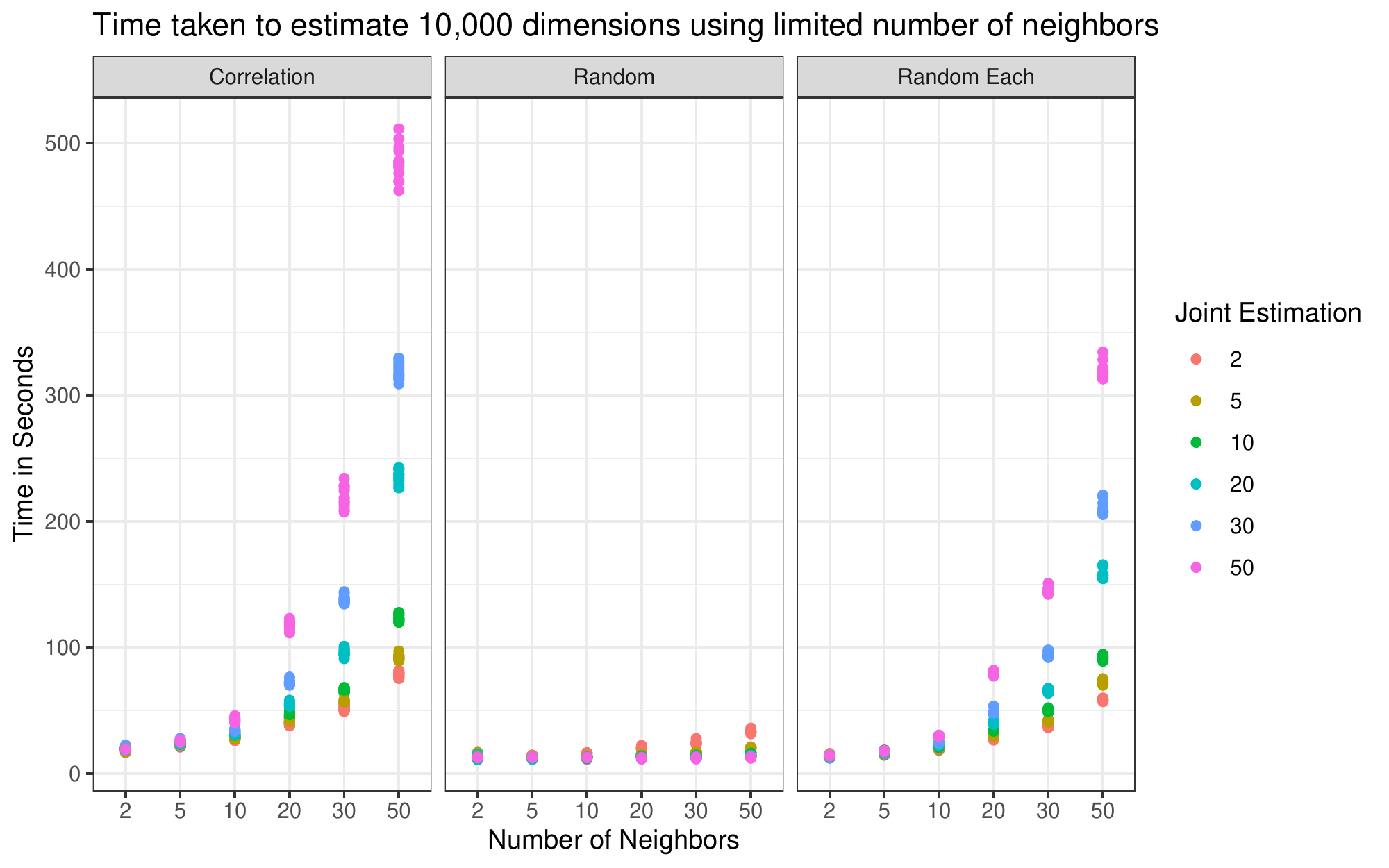}
    \caption{Time to estimate the log CDF with dependence parameter $\rho=1$ using 3 different methods to select neighbors, multiple number of neighbors and multiple number of joint observations.}
    \label{fig:TimeNeighborsJoint}
\end{figure}

Figures \ref{fig:LogCumLikNeighborsJoint} and \ref{fig:TimeNeighborsJoint} show the estimated log {\cdf} and time (in seconds) to compute the approximated log {\cdf}, using the $100\times 100$ grid.  We used approximations based on joint conditional {\cdf}s of dimension 2, 5, 10, 20, 30, and 50.  For each grouping size, we constructed conditioning sets using 3 different methods.  The first method conditions on the $m$ most correlated observations (in this case simply the nearest neighbors) for each observation in the joint grouping, resulting in a conditioning set of size $pm$.  The second method simply conditions on $m$ random observations.  The third method conditions on $m$ random observations per element of the multivariate conditional calculation, again resulting of a conditioning set of size $pm$.

From Figure \ref{fig:LogCumLikNeighborsJoint} it is clear that simply conditioning on $m$ random observations fails to yield an acceptable approximation. Performance can be improved by conditioning on more random observations, which is what the third method does. Method 3 shows somewhat improved behavior, however it was only able to perform acceptably when both the dimensionality $p$ of the joint conditional probability and the size of the conditioning set $pm$ were both large.  It is clear from Figure \ref{fig:LogCumLikNeighborsJoint} that conditioning on random neighbors is much less accurate than conditioning on the most highly correlated neighbors.  For conditioning sets consisting of small numbers $m$ of neighbors per element in the joint conditional probability, the use of a large group $p$ of joint observations had a better result, probably simply due to fact that the total number $pm$ of neighbors in the conditioning set was larger.  However, when the number $m$ of correlated neighbors gets large enough the number $p$ of joint observations does not seem to affect result of the approximation.

Figure \ref{fig:TimeNeighborsJoint} shows the computation time required for all of the approximation schemes depicted in Figure \ref{fig:LogCumLikNeighborsJoint}.  The clear trend is that choosing a small conditioning set of random observations is very fast (middle panel), using higher-dimensional joint conditional {\cdf}s is slower than using lower-dimensional joint conditional {\cdf}s (all panels), and for the same size conditioning set, the time required to find the nearest neighbors is not a major bottleneck (right and left panels).  This conclusion is different from exploration of the same issues, in the context of the {\pdf}, found in \citet{Guinness2016}.  There, using higher-dimensional joint conditional calculations was found to be beneficial, and the time required to find nearest neighbors was substantial enough to warrant the use of a fast approximate ordering algorithm.  In the case of the {\cdf} approximation, code profiling confirmed that the time required to order the observations was insignificant, with the overwhelming majority of the computation time being used in calculating the lower-dimensional joint {\cdf}s using the QM technique.

\section{Example: A Gaussian Scale Mixture for Spatial Extremes}
\label{sec:GSMM}

Recent advances in the statistics of extremal spatial phenomena have produced models that are flexible enough to accommodate both strong and weak spatial dependence in the far joint tails.  One prominent strategy for achieving this is to construct scale mixtures of Gaussian processes, where the mixing distribution is chosen carefully so as to produce the desired tail dependence characteristics \citep{Opitz2016,Huser2017,Morris2017,Huser2019}.  The preferred flavor of maximum likelihood inference for these models requires computing a Gaussian {\cdf} whose dimension is roughly equal to the number of spatial locations in the dataset.  Other state-of-the-art models for spatial extremes also rely on high-dimensional Gaussian {\cdf}s \citep{Wadsworth2014,Thibaud2016,deFondeville2018}.  To show the usefulness of our {\cdf} approximation, we analyze data from precipitation gauges in Europe using the Gaussian scale mixture model from \citet{Huser2017}, which we describe below. 

The class of scale mixtures of Gaussian processes is defined generically by 
\begin{align}
\label{eq:model01}
    X(\bs) & = R \times W(\bs) \nonumber \\ 
    R & \sim F_R \indep W(\bs). 
\end{align}
Here, $W(\bs)$ is a standard Gaussian process (i.e. with unit variance) on some domain $\mathcal{D}$ indexed by $\bs \in \mathcal{D}$.    For a collection of $k$ observations, the finite dimensional distribution of the Gaussian component is $\mathbf{W}\sim N_k(0,\bSigma(\theta))$, where $\bSigma(\theta)$ is a $D \times D$ covariance matrix constructed using a chosen covariance model that is indexed by parameter $\theta$.

The random scaling $R$ comes from distribution $F_R$.  The choice of $F_R$ is critical and determines the strength of the tail dependence in the resulting model \citep{Engelke2019}.  A key quantity for summarizing the strength of tail dependence is the conditional probability $\chi_u(\bs_i,\bs_j) = P\{X(\bs_i) > u \given X(\bs_j > u\}$, for spatial locations $\bs_i$ and $\bs_j$.  If $\lim_{u \rightarrow \infty}\chi_u(\bs_i, \bs_j) = 0$ for all $\bs_i, \bs_j \in \mathcal{D}$, we say that $X(\bs)$ is \emph{asymptotically independent}, while if $\lim_{u \rightarrow \infty}\chi_u(\bs_i, \bs_j) > 0$ for all $\bs_i, \bs_j \in \mathcal{D}$, we say that $X(\bs)$ is \emph{asymptotically dependent}.

While many choices are available for the mixing distribution $F_R$, \citet{Huser2017} suggest the parametric model defined by equation \eqref{eq:model02}. When $\beta>0$, the mixture process $X(\bs)$ is asymptotically independent, and when $\beta=0$, $X(\bs)$ is asymptotically dependent.  Therefore, this class of scale mixtures is rich enough to include both asymptotic independence and asymptotic dependence as nontrivial sub-models.

\begin{align}
\label{eq:model02}
    F_R(r) = \begin{cases} 1-\exp\{ -\gamma(r^{\beta}-1)/\beta \}, & \text{for } \beta>0 \\
    1-r^{\gamma}, & \text{for } \beta=0. \end{cases}
\end{align}

To construct the likelihood for maximum likelihood estimation, we must integrate out $R$ from the model \eqref{eq:model01}. Equations \eqref{eq:model03}  and \eqref{eq:model03-pdf} show the marginal multivariate {\cdf} and {\pdf}, respectively, for a finite collection of observations $\mathbf{X}$ from $X(\bs)$ defined in \eqref{eq:model01}. Here $\Phi_D$ represents the $D$-dimensional multivariate {\cdf} from a Gaussian distribution with mean vector $0$ and covariance matrix $\bSigma(\theta)$, and $\phi_D$ represents the $D$-dimensional multivariate {\pdf} from a Gaussian distribution with mean $0$ and covariance matrix $\bSigma(\theta)$. There are no closed forms for these expressions, so it is necessary to use numerical methods to evaluate the (one-dimensional) integrals.

\begin{align}
\label{eq:model03}
    G(\mathbf{x}) & =\int_0^\infty \Phi_D(\mathbf{x}/r;\boldsymbol{\Sigma})f_R(r)dr\\
    \label{eq:model03-pdf}
    g(\mathbf{x}) & =\int_0^\infty \phi_D(\mathbf{x}/r;\boldsymbol{\Sigma})r^{-D}f_R(r)dr.
\end{align}

The preferred strategy for maximum likelihood estimation of extremal dependence models is to treat all observations falling below a high threshold as left censored \citep{Huser2016b}.  This leads to a favorable balance between using the data as efficiently as possible, while not allowing data in the bulk of the distribution to have a large effect on dependence estimation.  The censored likelihood for each temporal replicate is obtained by taking one partial derivative of \eqref{eq:model03} for every observation that falls above the threshold.  Thus,  \eqref{eq:model03-pdf} is the relevant likelihood when all observations, at one particular temporal replicate, are above the threshold, so nothing is censored. However, since the threshold is chosen to be a high quantile to prioritize inference on the tail, most observations are usually censored for any temporal replicate.  When all observations fall below the threshold, the relevant likelihood is \eqref{eq:model03}.

Most often, in any temporal replicate, there will be a mixture of observations above and below the threshold. In this case, the relevant joint likelihood of $\bx$ is defined by equation \eqref{eq:model05}, which results from taking partial derivatives of \eqref{eq:model03} with respect to only the un-censored observations. If we let $I$ be the set of points above the threshold and $I^c$ be the points below, then

\begin{align}
\label{eq:model05}
    G_I(\bx) & \coloneqq \frac{\partial^{\mid I \mid}}{\partial \bx_I}G(\bx)=\int_0^\infty \frac{\partial^{\mid I \mid}}{\partial \bx_I}\Phi_k(\bx/r;\boldsymbol{\bSigma})f_R(r)dr \nonumber \\
    & =\int_0^\infty \Phi_{\mid I^c \mid}\{(\bx_{I^c}-\bSigma_{I^c;I}\bSigma^{-1}_{I;I}\bx_I)/r;\bSigma_{I^c\mid I} \}\phi_{\mid I \mid}(\bx_I/r;\bSigma_{I;I})r^{-\mid I \mid} f_R(r)dr,
\end{align}
where dependence of the covariance matrices on $\theta$ is suppressed for brevity, and the notation  $\bSigma_{A;A}$ refers to rows and columns of $\bSigma$ pertinent to the points in $A$.  The matrix $\bSigma_{I^c\mid I}=\bSigma_{I^c;I^c}-\bSigma{I^c;I}\bSigma_{I;I}^{-1}\bSigma_{I;I^c}$ is the covariance matrix of the conditional normal distribution of the censored observations given the un-censored observations.

The computational issue arises because the integrand \eqref{eq:model05} contains a Gaussian {\cdf} of dimension $|I^c|$, the number of censored observations in a temporal replicate.  Again, for most replicates, this number $|I^c|$ is close to the total number of observation locations $D$ because the censoring threshold is chosen to be high, such that most observations fall below the threshold and are therefore censored.

\subsection{Precipitation Over Europe}%Plots
\label{subsec:Data}

\begin{figure}[H]
    \centering
    \includegraphics[scale=0.9]{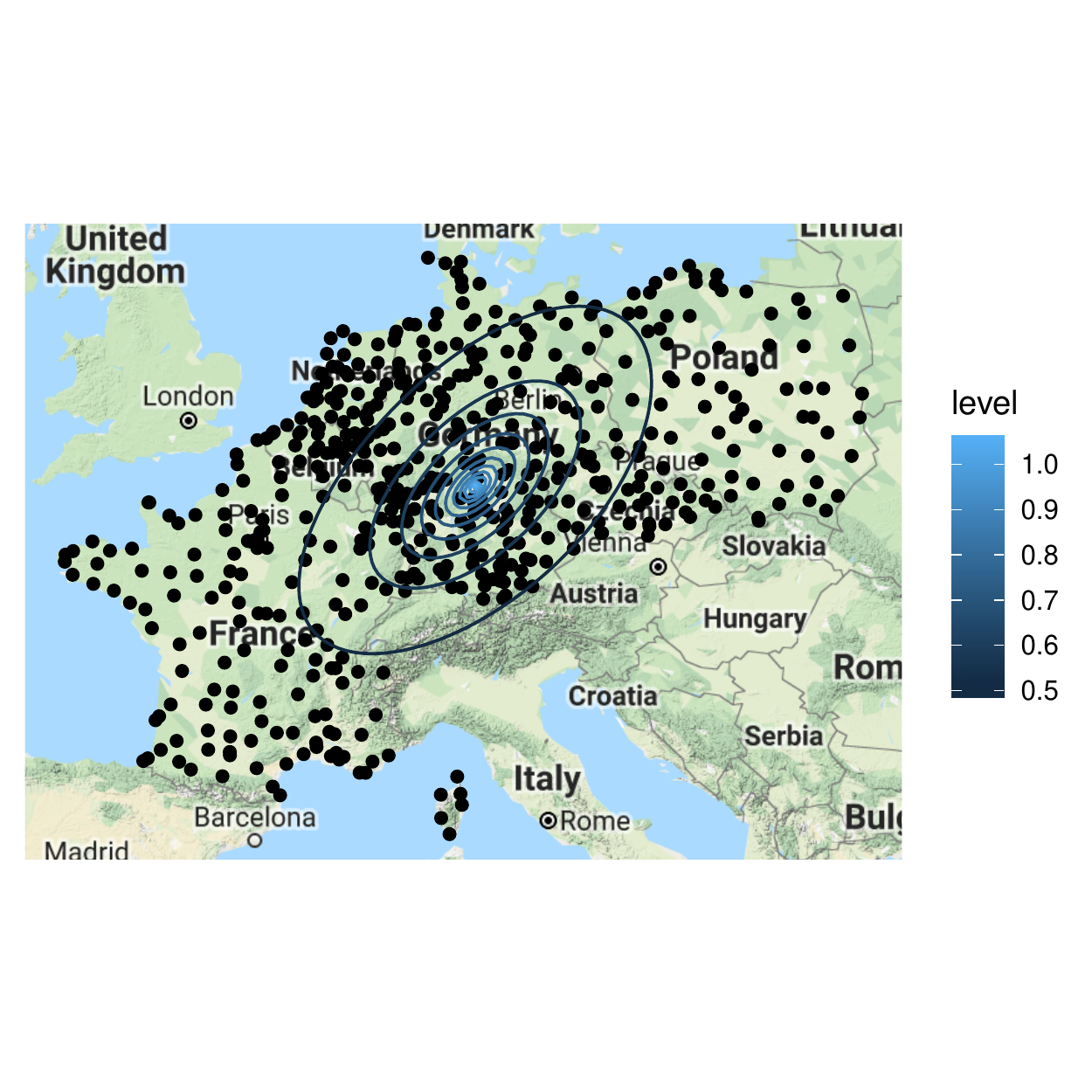}
    \caption{$D=528$ weather stations located over 6 European countries}
    \label{fig:StationLoc}
\end{figure}

Our dataset consists of weekly maximum precipitation observations between January, 2000 and April, 2019, in the western and central region of continental Europe, north of the mountain ranges the Pyrenees, Alps, and Carpathians.  The 6 countries we consider are Germany, Poland, Netherlands, Belgium, Czech Republic, and France.  Figure \ref{fig:StationLoc} shows the locations of the observation stations distributed over Europe.  This dataset consists of 1,006 weekly maxima from $D=528$ weather stations.  For context, the computational bottleneck from the Gaussian {\cdf} limited the analysis in \citet{Huser2017} to a dataset of $D=12$ locations, even though analysis was performed on a large high-performance computing cluster.  We use the weekly maximum daily accumulations at each location to break temporal dependence that might arise from storms that persist for more than one day.  Out of the 531,168 total observations, 32.6\% were missing values. For each weekly maximum, only the available data was used for estimation, and all missing observations were disregarded.

The covariance model we use for the underlying Gaussian processes is an anisotropic exponential, $\Sigma_{ij}(\theta) = \exp\{-h_{ij}/\rho\}$, where $\rho$ is the range parameter and $h_{ij}$ is the Mahalanobis distance between locations $\bs_i$ and $\bs_j$.  The Mahalanois distance is parametrized as 
\[
  h_{ij}^2 = \Omega \trans \Omega, \quad \text{where} \, 
  \Omega = (\bs_i - \bs_j)\trans 
    \begin{pmatrix}
      \cos(\phi) & -\sin(\phi) \\
      \sin(\phi) & \cos(\phi) \\
    \end{pmatrix}^{-1}
    \begin{pmatrix}
      1 & 0 \\
      0 & A \\
    \end{pmatrix},
\]
for rotation angle $\phi \in [0, \pi)$ and and aspect ratio $A > 1$. Thus,  after fixing the mixing parameter $\gamma$ at 1, as it plays a much less significant role than the parameter $\beta$ in determining tail dependence characteristics, we arrive at a total of 4 parameters to estimate, $\psi=( \beta, \rho, \phi, A)\trans$.

The first step in estimating the dependence is to transform the observations to be on the same marginal scale.  To do this, we start by applying a rank transformation to standard uniform, independently for each station.  That is, for each station $k=1, \ldots, D$ and each time point  $t=1, \ldots T$, the observation $X_{kt}$ on the uniform scale is
\begin{align*}
% \label{eq:RankTrans}
    U_{kt} = \frac{\rank(X_{kt})}{T+1}.
\end{align*}
We next choose a high threshold to be the 0.95 marginal empirical quantile at each location.  Then, denoting the marginal {\cdf} and {\pdf} of each $X_{kt}$, respectively, as $G_M({x})=\int_0^\infty \Phi({x}/r)f(r)dr$ and $g_M(x)=\int_0^\infty \phi({x}/r)r^{-1}f(r)dr$ (we assume stationarity, so the marginal distribution is assumed to be the same at each location), and letting the vector $\bv_{t} = (\max\{u_{1t}, 0.95\}, \ldots, \max\{u_{Dt}, 0.95\})\trans$, the copula censored likelihood for each time replicate $k$ is
\begin{align*}
    % \label{eq:censored-copula-likelihood}
    L(\psi; \bv_t) = \begin{cases}
                       G\{G_M^{-1}(v_{1t}), \ldots, G_M^{-1}(v_{Dt})\} & \,\text{if all obs. are below the threshold} \\
                       \frac{g\{G_M^{-1}(v_{1t}), \ldots, G_M^{-1}(v_{Dt})\}}{\prod_{k=1}^D g_M\{G_M^{-1}(v_{kt})\}} & \,\text{if all obs. are above the threshold} \\
                       \frac{G_{I_t}\{G_M^{-1}(v_{1t}), \ldots, G_M^{-1}(v_{Dt})\}}{\prod_{k\in I_i}^D g_m\{G_M^{-1}(v_{kt})\}}  & \,\text{if some obs. are above and some below the threshold} 
                     \end{cases}
\end{align*}

% is transformed following equation \ref{eq:RankTrans}. A new vector of observations is than defined by $\mathbf{u}_i^*=\{max(u_{1,i},v_1),...max(u_{D,i},v_D) \}$ where $\mathbf{v}$ is the vector with threshold in uniform scale.

% The marginal distribution of a single $X_k$ can be estimated by considering only the marginal distribution of the D-dimensional Gaussian distribution. Equation \ref{eq:model04} represents the marginal distribution of $X_k$.

% \begin{align}
% \label{eq:model04}
%     G_k({x_k})=\int_0^\infty \Phi({x_k}/r)f(r)dr\\
%     g_k({x_k})=\int_0^\infty \phi({x_k}/r)r^{-1}f(r)dr
% \end{align}

% The likelihood for each time point, $i$, is then defined as 

% \begin{itemize}
%     \item if all observations are below the threshold
%     \begin{align*}
%         L(\mathbf{u}_i)=G(G_1^{-1}(v_1),...,G_D^{-1}(v_D))
%     \end{align*}
%     \item if all observations are above the threshold
%     \begin{align*}
%         L(\mathbf{u}_i)=\frac{g(G_1^{-1}(u_{1i}),...,G_D^{-1}(u_{Di}))}{\prod_{k=1}^D g_k(G_k^{-1}(u_{ki}))}
%     \end{align*}
%     \item if there is a mixture of observations above and below the threshold
%     \begin{align*}
%         L(\mathbf{u}_i)=\frac{G_{I_i}(G_1^{-1}(u^*_{1i}),...,G_D^{-1}(u^*_{Di}))}{\prod_{k\in I_i}^D g_k(G_k^{-1}(u_{ki}))}
%     \end{align*}
% \end{itemize}

Finally, the log likelihood across all time points $t$ for the parameter vector $\psi$ is 
\begin{align*}
% \label{eq:logLik}
        l(\psi; \bv)= \sum_{t=1}^T \log(L(\psi; \bv_t)).
\end{align*}

 We found the maximum likelihood estimator (MLE) by applying the Nelder-Mead numerical optimizer in the \texttt{R} function \texttt{optim}.  MLEs are shown in Table \ref{tab:par}.  The MLE for the mixing parameter $\beta$ is 0.82, which in this context is fairly far away from zero---far enough to strongly suggest that the process is asymptotically independent.  The MLEs for the anisotropy parameters suggest pronounced eccentricity.  To interpret and visualize the estimated dependence model implied by the MLEs shown in Table \ref{tab:par}, we plot level curves in the resulting $\chi_u$ function for $u=0.95$ on the quantile scale, shown in Figure \ref{fig:StationLoc}.  Each ellipse represents a constant value of $\chi_{u=0.95}(\bs) = P\{F_M[X(\bs)] > 0.95 \given F_M[X(\bs_0)] > 0.95 \}$, for an arbitrarily-chosen reference point $\bs_0$ near the center of the map.  The level curves are ellipses due to the anisotropic construction, with the major axis roughly along a northeast-southwest orientation, and joint exceedances more likely with decreasing distance from $\bs_0$.
 
%  For the same set of values the value of the log likelihood can be different since we use a QM step inside our function. So in order to access convergence we observed the log likelihood evolution in figure \ref{fig:logLik}. We can observe that after 20 interactions the log likelihood as stabilized. Figure \ref{fig:parEvol} show us that the same occurred with the parameters, they have stabilized in the end of the chain. Based on this chain table \ref{tab:par} reports the estimated value of the parameter in the last interaction.

% \begin{figure}[H]
%     \centering
%     \includegraphics[scale=0.4]{figure/Data/dens2.pdf}
%     \caption{Log Likelihood evolution}
%     \label{fig:logLik}
% \end{figure}

% \begin{figure}[H]
%     \centering
%     \includegraphics[scale=0.4]{figure/Data/par2.pdf}
%     \caption{Parameter evolution}
%     \label{fig:parEvol}
% \end{figure}

\begin{table}[H]
\centering
\begin{tabular}{lr}
\hline
Parameter & \multicolumn{1}{l}{MLE} \\ \hline
$\rho$     & 1.31    \\
$\beta$    & 0.82    \\
$\phi$     & 1.10    \\  % $1.10 = 0.35\pi$
$A$    & 2.29        \\ \hline
\end{tabular}
\caption{Maximum likelihood estimates of dependence parameters}
\label{tab:par}
\end{table}

\section{Discussion}

The main objective of this paper was to propose fast approximation to high-dimensional Gaussian {\cdf}s that arise from spatial Gaussian processes.  We modified Vecchia's approximation for Gaussian {\pdf}s to the context of Gaussian {\cdf}s.  Simulations showed that for large numbers of locations and relatively small conditioning sets, this approximation gives results consistent with state-of-the-art QM methods, and reduces computational time considerably, even when computations are performed sequentially.  Furthermore, the approximation is trivially easy to code in parallel using standard \texttt{R} packages, and requiring no linking to specialized software libraries.

We demonstrated the utility of our fast {\cdf} approximation by using it to find maximum censored likelihood estimates for the scale mixture model of \citet{Huser2017}.  This model is attractive because of its flexible tail dependence characteristics, but is hampered by computational difficulties arising from the need to compute high-dimensional Gaussian {\cdf}s during inference.  We fit this model to a precipitation dataset consisting over 500 spatial locations, whereas previous efforts using conventional QM techniques were limited to just 12 locations.

One drawback that we noticed during the data analysis is that conventional optimization routines had trouble converging, due to the stochastic nature of the likelihood objective function.  For future studies, one possible approach to circumventing this problem is to use stochastic optimization algorithms, which may be better suited to optimizing  random objective functions.

\section*{Acknowledgements}

This research was supported in part by NSF grant DMS-1752280. Computations for this research were performed on the Institute for Computational and Data Sciences Advanced CyberInfrastructure (ICDS-ACI).

\bibliographystyle{plainnat}
\bibliography{vecchia.bib}

\end{document}